%% file: NCPBP.tex
\documentclass[conference]{IEEEtran}

\IEEEsettopmargin{t}{0.755in}

\usepackage{cite}
\usepackage{graphicx}
\usepackage{subcaption}
\usepackage{xcolor}
\usepackage{pgfplots}
\usepackage{multirow}
\usepackage{upgreek}
\usepackage{amssymb}
\usepackage{amsmath}
\usepackage{cases}
\usepackage{pifont}
\usepackage{bm}
\usepackage{amsthm}

\usepackage{tabularx}
\usepackage{booktabs}
\newcolumntype{Y}{>{\centering\arraybackslash}X}
\usepackage{array}
\usepackage{float}
\usepackage{xcolor}
\usepackage[linesnumbered,ruled,vlined]{algorithm2e}
\usepackage{algorithmic}
\usepackage{changes}
\usepackage{bbm}
\usetikzlibrary{arrows,positioning,shapes.geometric,spy}

\newcommand{\fixme}[2]{\ifx&#2&{\leavevmode\color{red}#1}\else{\leavevmode\color{red}FIXME\{}#1{\leavevmode\color{red}\}}\footnote{{\leavevmode\color{red}#2}}\PackageWarning{Fixme}{#1: #2}\fi}

\SetCommentSty{mycommfont}

\DeclareMathOperator*{\sgn}{sgn}

\DeclareMathOperator*{\arctanh}{arctanh}

\begin{document}
	
	\title{Neural Belief Propagation Decoding of \\ CRC-Polar Concatenated Codes}
	
	\author{\IEEEauthorblockN{Nghia Doan, Seyyed Ali Hashemi, Elie Ngomseu Mambou, Thibaud Tonnellier, Warren J. Gross}
		\IEEEauthorblockA{Department of Electrical and Computer Engineering, McGill University, Montr\'eal, Qu\'ebec, Canada\\
			Email: \{nghia.doan, seyyed.hashemi, elie.ngomseumambou\}@mail.mcgill.ca, \{thibaud.tonnellier, warren.gross\}@mcgill.ca}}
	
	\maketitle
	\begin{abstract}
		Polar codes are the first class of error correcting codes that provably achieve the channel capacity at infinite code length. They were selected for use in the fifth generation of cellular mobile communications (5G). In practical scenarios such as 5G, a cyclic redundancy check (CRC) is concatenated with polar codes to improve their finite length performance. This is mostly beneficial for sequential successive-cancellation list decoders. However, for parallel iterative belief propagation (BP) decoders, CRC is only used as an early stopping criterion with incremental error-correction performance improvement. In this paper, we first propose a CRC-polar BP (CPBP) decoder by exchanging the extrinsic information between the factor graph of the polar code and that of the CRC. We then propose a neural CPBP (NCPBP) algorithm which improves the CPBP decoder by introducing trainable normalizing weights on the concatenated factor graph. Our results on a 5G polar code of length $128$ show that at the frame error rate of $10^{-5}$ and with a maximum of $30$ iterations, the error-correction performance of CPBP and NCPBP are approximately $0.25$~dB and $0.5$~dB better than that of the conventional CRC-aided BP decoder, respectively, while introducing almost no latency overhead.
	\end{abstract}
	
	\IEEEpeerreviewmaketitle
	\section{Introduction} \label{sec:intro}
	
	Polar codes are a breakthrough in the field of channel coding as they were proved to achieve channel capacity with efficient encoding and decoding algorithms \cite{arikan}. Successive cancellation (SC) and belief propagation (BP) decoding algorithms are introduced in \cite{arikan} to decode polar codes. Although SC decoding can provide a low-complexity implementation, its serial nature prevents the decoder to reach a high decoding throughput. In addition, the error-correction performance of SC decoding for short to moderate polar codes does not satisfy the requirements of the fifth generation of cellular mobile communications (5G). To improve the performance of SC decoding, SC list (SCL) decoding was introduced in \cite{tal_list} and it was shown that SCL can provide a vast error-correction performance improvement if it is aided by a cyclic redundancy check (CRC). Based on this observation, polar codes have been selected to be used in the enhanced mobile broadband (eMBB) control channel of 5G together with a CRC \cite{3gpp_report}.
	
	Unlike SC-based decoders, the iterative message passing process of BP decoding can be executed in parallel, hence enabling the decoder to reach high decoding throughput. However, with limited number of iterations BP decoding suffers from poor error-correction performance. Furthermore, 5G standard requires polar codes to be concatenated with an outer CRC code. Thus, several attempts have been carried out to improve the performance of BP decoding for polar codes using a CRC. In \cite{RenBPTerminate}, CRC is used as an early termination criterion to prevent the BP decoder from processing unnecessary iterations when the correct codeword is found. In \cite{Sun}, a post-processing algorithm is presented which uses a CRC to detect false-converged errors. In addition, it is observed in \cite{elkelesh2018belief,Doan_GLOBECOM} that BP decoding on a list of factor graph permutations of polar codes can benefit from CRC to achieve lower error probabilities than when no CRC is used. However, BP decoding in all of the aforementioned works is only applied on the factor graph of polar codes and the CRC is only used to verify the result of BP at each iteration, without exploiting the inherent factor graph of CRC. Note that BP decoding is used in \cite{Guo, Abbas, Elkelesh} on the concatenated factor graphs of a low-density parity-check (LDPC) code and a polar code. However, this scheme is not selected for use in the eMBB control channel of 5G. 
		
	In this paper, we first show that by running BP decoding on the CRC-polar concatenated factor graph, significant error-correction performance improvement can be achieved in comparison with the conventional CRC-aided BP decoder. We call the proposed decoding method CRC-polar BP (CPBP). We then propose a neural CPBP (NCPBP) decoder to further improve the error-correction performance of CPBP with limited number of BP iterations. We devise an efficient weight-assignment scheme for NCPBP and show that the proposed scheme has fewer weights than the state-of-the-art neural BP decoder of \cite{Nachmani_STSP} while providing a better error-correction performance. The proposed decoders are evaluated on a 5G polar code of length $128$. At the frame error rate (FER) of $10^{-5}$ and with a maximum number of $30$ iterations, we show that the error-correction performance of the CPBP and NCPBP decoders are approximately $0.25$~dB and $0.5$~dB better than that of the conventional CRC-aided BP decoder, respectively, with negligible latency overhead.
	
	The rest of this paper is organized as follows. Section~\ref{sec:polar} briefly introduces polar codes and its BP-based decoders. Section~\ref{sec:CPBP} and \ref{sec:NCPBP} describe the proposed CPBP and NCPBP decoders, respectively. Finally, concluding remarks are drawn in Section~\ref{sec:conclude}. 
	
	\section{Preliminaries} \label{sec:polar}
		
	\subsection{Polar Codes}
	\label{sec:polar:polar} 
	
	A polar code $\mathcal{P}(N,K)$ of length $N$ with $K$ information bits is constructed by applying a linear transformation to the message word $\bm{u} = \{u_0,u_1,\ldots,u_{N-1}\}$ as $\bm{x} = \bm{u}\bm{G}^{\otimes n}$ where $\bm{x} = \{x_0,x_1,\ldots,x_{N-1}\}$ is the codeword, $\bm{G}^{\otimes n}$ is the $n$-th Kronecker power of the polarizing matrix $\bm{G}=\bigl[\begin{smallmatrix} 1&0\\ 1&1 \end{smallmatrix} \bigr]$, and $n = \log_2 N$. The vector $\bm{u}$ contains a set $\mathcal{A}$ of $K$ information bits and a set $\mathcal{A}^c$ of $N-K$ frozen bits. The positions of the frozen bits are known to the encoder and the decoder and their values are known to be $0$. The codeword $\bm{x}$ is then modulated and sent through the channel. In this paper, binary phase-shift keying (BPSK) modulation and additive white Gaussian noise (AWGN) channel model are considered, therefore, the soft vector of the transmitted codeword received by the decoder is written as
	\begin{equation}
	\bm{y}=(\mathbf{1}-2\bm{x})+\bm{z}\text{,}
	\end{equation}
	where $\mathbf{1}$ is an all-one vector of size $N$, and $\bm{z} \in \mathbbm{R}^N$ is the AWGN noise vector with variance $\sigma^2$ and zero mean. In the log-likelihood ratio (LLR) domain, the LLR vector of the transmitted codeword is
	\begin{equation}
\mathbf{LLR}_{\bm{x}} = \ln{\frac{\text{Pr}(\bm{x}=0|\bm{y})}{\text{Pr}(\bm{x}=1|\bm{y})}}=\frac{2\bm{y}}{\sigma^2}\text{.}
	\end{equation}
	
	\subsection{Belief Propagation Decoding of Polar Codes}
	\label{sec:polar:BPD}
		
	\begin{figure}[t]
		\vspace*{0.3\baselineskip}
		\centering
		\begin{subfigure}[b]{0.5\textwidth}
			\centering
			\includegraphics[width=0.6\linewidth]{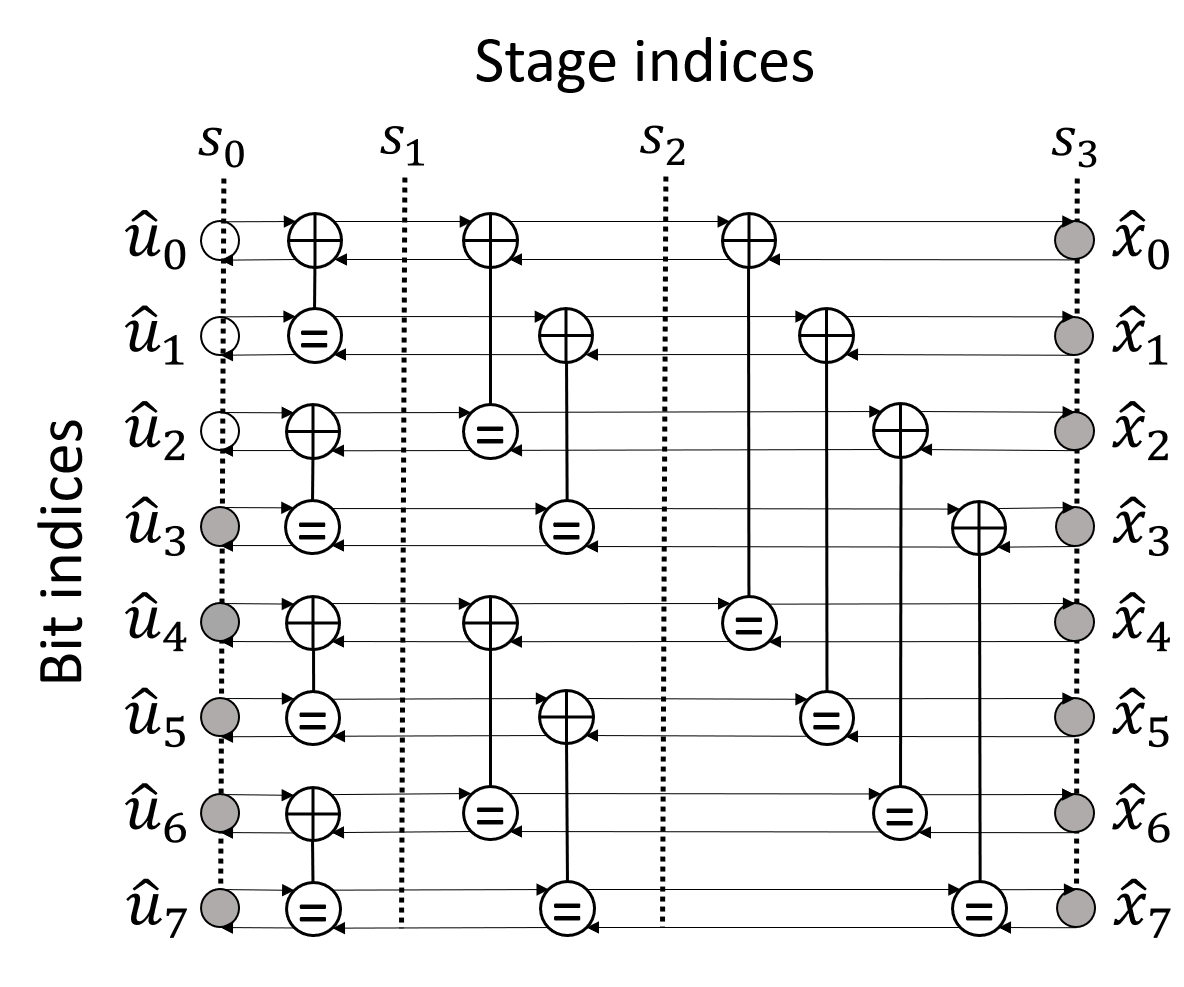}
			\caption{}
			\label{figBPDec:a}
		\end{subfigure} \\   
		\begin{subfigure}[b]{0.3\linewidth}
			\centering
			\includegraphics[width=0.8\linewidth]{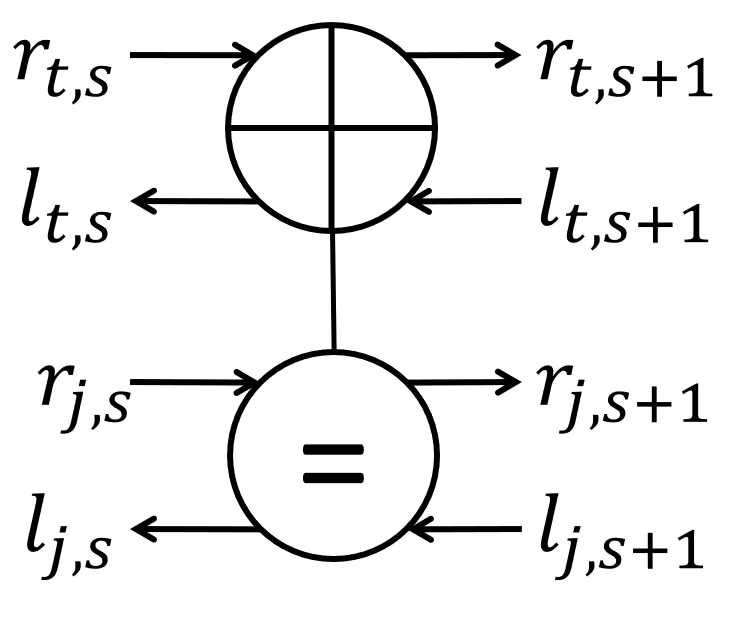}
			\caption{}
			\label{figBPDec:b}
		\end{subfigure}  
		\begin{subfigure}[b]{0.3\linewidth}
			\centering
			\includegraphics[width=0.8\linewidth]{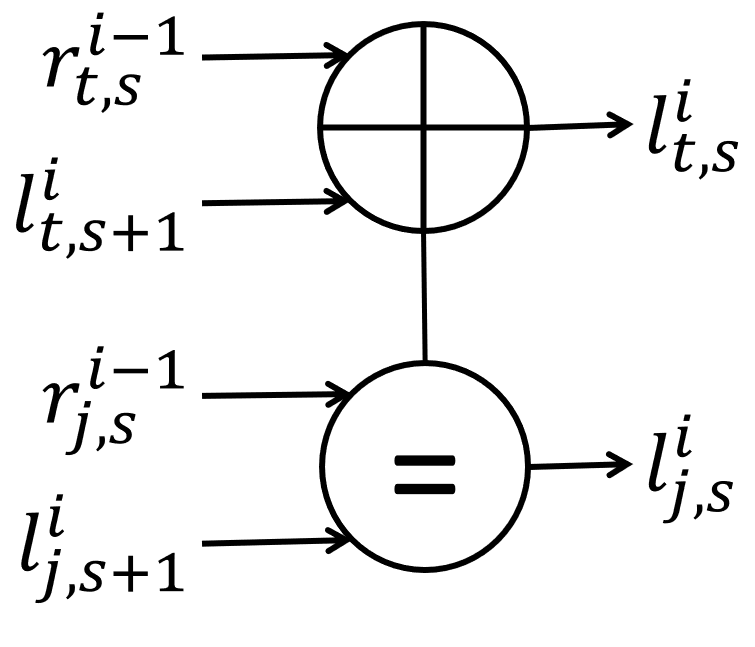}
			\caption{}
			\label{figBPDec:c}
		\end{subfigure}  
		\begin{subfigure}[b]{0.3\linewidth}
			\centering
			\includegraphics[width=0.9\linewidth]{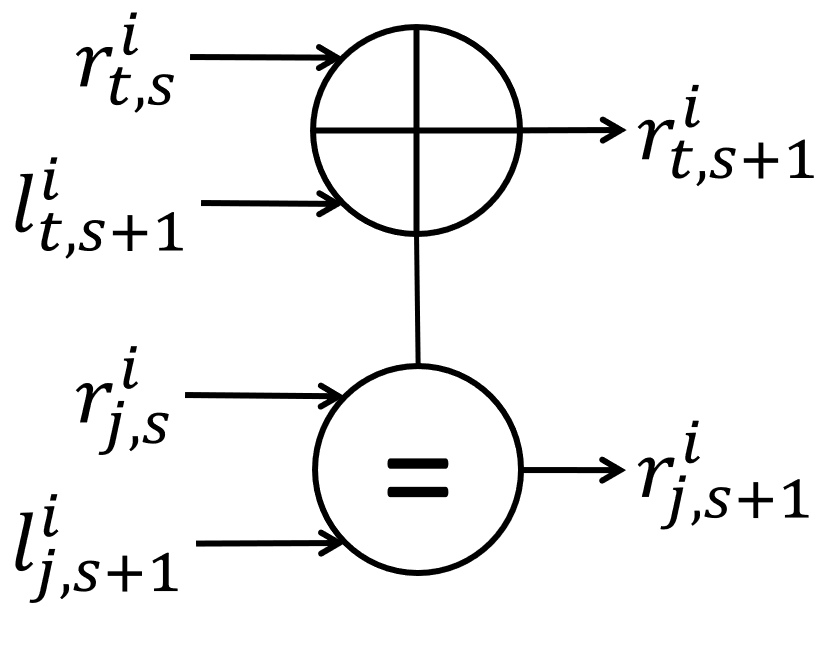}
			\caption{}
			\label{figBPDec:d}
		\end{subfigure}  
		\caption{(a) BP decoding on the factor graph of $\mathcal{P}(8,5)$ with $\{u_0,u_1,u_2\}\in \mathcal{A}^c$, (b) a PE, (c) a right-to-left message update of a PE on an unrolled factor graph, and (d) a left-to-right message update of a PE on an unrolled factor graph.}
		\vspace*{-1\baselineskip}
	\end{figure}		
		
	Fig.~\ref{figBPDec:a} illustrates BP decoding on a factor graph representation of $\mathcal{P}(8,5)$. The messages are iteratively propagated through the processing elements (PEs) \cite{arikan2010polar} located in each stage. An update iteration starts with a right-to-left message pass that propagates the LLR values from the channel (rightmost) stage, to the information bit (leftmost) stage, and ends with the left-to-right message pass which occurs in the reverse order. Fig.~\ref{figBPDec:b} shows a PE with its corresponding messages, where $r_{t,s}$ denotes a left-to-right message, and $l_{t,s}$ denotes a right-to-left message of the $t$-th bit index at stage $s$. Equivalently, BP decoding of polar codes can be represented on an unrolled factor graph, in which BP iterations are performed sequentially. Fig.~\ref{figBPDec:c} and Fig.~\ref{figBPDec:d} illustrate the input and output messages of a PE for the right-to-left and left-to-right message updates on an unrolled factor graph, where the superscript $i$ denotes the iteration number. The update rule \cite{arikan2010polar} for the right-to-left messages of a PE is
	\begin{align}
	\label{PolarPE_left}
	\begin{split}
	\begin{cases}
	l^{i}_{t,s} &= f(l^{i}_{t,k},r^{i-1}_{j,s} + l^{i}_{j,k})\text{,}\\
	l^{i}_{j,s} &= f(l^{i}_{t,k},r^{i-1}_{t,s}) + l^{i}_{j,k}\text{,}\\
	\end{cases}
	\end{split}
	\end{align}
and for the left-to-right messages is
	\begin{align}
	\label{PolarPE_right}
	\begin{split}
	\begin{cases}
	r^{i}_{t,k} &= f(r^{i}_{t,s},l^{i}_{j,k} + r^{i}_{j,s})\text{,}\\
	r^{i}_{j,k} &= f(r^{i}_{t,s},l^{i}_{t,k}) + r^{i}_{j,s}\text{,}
	\end{cases}
	\end{split}
	\end{align}
where $j=t+2^s$, $k=s+1$, and
	\begin{equation}
	\label{SPA}
	f(x,y)=2\arctanh \left( \tanh{\left(\frac{x}{2}\right)}\tanh{\left(\frac{y}{2}\right)} \right) \text{,}
	\end{equation}
for any $x,y \in \mathbbm{R}$. Note that implementing (\ref{SPA}) is costly in practice, instead the following approximation of (\ref{SPA}) is used in this paper:
	\begin{equation}
	\label{minsum}
	f(x,y) \approx \tilde{f}(x,y) = \sgn(x)\sgn(y)\min(|x|,|y|)\text{.}
	\end{equation}
	
	The BP decoding performs a predetermined $I_{\max}$ update iterations where the messages are propagated through all PEs in accordance with (\ref{PolarPE_left}) and (\ref{PolarPE_right}). Initially, for $0 \leq t < N$ and $\forall i \le I_{\max}$, $l^{i}_{t,n}$ are set to the received channel LLR values $\mathbf{LLR}_{\bm{x}}$, $r^{i}_{t,0}$ are set to the LLR values of the information and frozen bits as
	\begin{equation}
	\mathbf{LLR}_{\mathcal{A} \cup \mathcal{A}^c} =
	\begin{cases}
	0 \text{,} & \text{if } t \in \mathcal{A} \text{,}\\
	+\infty \text{,} & \text{if } t \in \mathcal{A}^c \text{.}
	\end{cases}
	\end{equation}		
	All the other left-to-right and right-to-left messages of the PEs at the first iteration are set to $0$. After running $I_{\max}$ iterations, the decoder makes a hard decision on the LLR values of the $t$-th bit at the information bit stage to obtain the estimated message word as
	\begin{equation}
	\label{hardDec}
	\hat{u}_t=
	\begin{cases}
	0 \text{,} & \text{if } r^{I_{\max}}_{t,0} + l^{I_{\max}}_{t,0} \geq 0 \text{,}\\
	1 \text{,} & \text{otherwise.}
	\end{cases} 
	\end{equation}	
In the following sections the vector forms of the left-to-right and right-to-left messages at the $s$-th stage and the $i$-th iteration are denoted as $\bm{l}^i_s$ and $\bm{r}^i_s$, respectively, while that of the estimated message word is denoted as $\bm{\hat{u}}$.
		
A CRC is used for BP decoding to either early terminate the BP process \cite{RenBPTerminate}, or to help select the correct codeword among a list of candidates \cite{elkelesh2018belief,Doan_GLOBECOM}. However, these CRC utilizations do not take into account the factor graph realization of CRC, on which the BP decoder can be applied.
	
	\subsection{Neural BP Decoding}
	
	Neural BP decoding was introduced in \cite{Nachmani_STSP,Loren_ISIT_2017} to improve the error-correction performance of BP decoding on Bose-Chaudhuri-Hocquenghem (BCH) codes by assigning trainable weights to the conventional BP decoding. Neural normalized min-sum recurrent neural network (NNMS-RNN) is a powerful variant of neural BP \cite{Nachmani_STSP} with the following weight assignment scheme for the message update rule of a PE in (\ref{PolarPE_left}) and (\ref{PolarPE_right}):
	\begin{align}
	\label{PE_NNMSRNN_left}
	\begin{split}
	\begin{cases}
	l^{i}_{t,s} &= w_0 \tilde{f}(l^{i}_{t,k},w_1 r^{i-1}_{j,s} + w_2 l^{i}_{j,k})\text{,}\\
	l^{i}_{j,s} &= w_4(w_3\tilde{f}(l^{i}_{t,k},r^{i-1}_{t,s})) + w_5 l^{i}_{j,k}\text{,}\\
	\end{cases}
	\end{split}
	\end{align}
	\begin{align}
	\label{PE_NNMSRNN_right}
	\begin{split}
	\begin{cases}
	r^{i}_{t,k} &= w_6 \tilde{f}(r^{i}_{t,s},w_7 l^{i}_{j,k} + w_8 r^{i}_{j,s})\text{,}\\
	r^{i}_{j,k} &= w_{10} (w_9 \tilde{f}(r^{i}_{t,s},l^{i}_{t,k})) + w_{11} r^{i}_{j,s}\text{,}
	\end{cases}
	\end{split}
	\end{align}
	where $w_m \in \mathbbm{R}$ $(0 \leq m \leq 11)$ are the trainable weights.

	The NNMS-RNN BP decoder suffers from a large number of weights which adversely affects its implementation cost. A neural normalized min-sum (NNMS) decoder was used to decode polar codes by only enabling the training for $w_0, w_3, w_6$ and $w_9$, while setting the other weights in (\ref{PE_NNMSRNN_left}) and (\ref{PE_NNMSRNN_right}) to $1$ \cite{Xu}. However, the error-correction performance improvement of \cite{Xu} with respect to the conventional BP is incremental.
		
	\section{CRC-Polar BP Decoding}	
	\label{sec:CPBP}

\begin{figure}[t]
		\centering
		\includegraphics[width=0.7\linewidth]{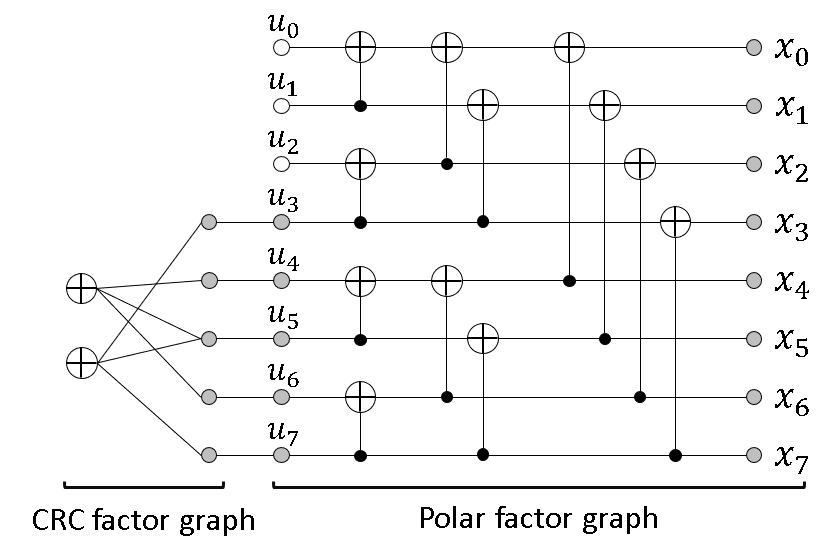}
		\caption{Factor graph representation of a CRC-polar concatenated code. The polar code is $\mathcal{P}(8,3)$ and a $2$-bit CRC is used.}
		\label{figBPCRC}	
		\vspace*{-1\baselineskip}
	\end{figure}

	In this section, we present the CPBP decoding algorithm which exploits the concatenated factor graph of a polar code and a CRC. Fig.~\ref{figBPCRC} shows the concatenated factor graph of $\mathcal{P}(8,3)$ and a CRC of length $2$. We run BP decoding algorithm on the concatenated factor graph to exploit the extrinsic information of the two constituent factor graphs. A similar approach was performed in \cite{Guo} for a LDPC-polar concatenated code by passing the BP messages between the factor graphs of LDPC and polar code at each iteration. A direct application of the BP decoder in \cite{Guo} to the CRC-polar concatenated code is not beneficial. This is due to the fact that the LDPC code is only connected to a few pre-selected information bits of polar codes which ensures the extrinsic information received by polar code is reliable enough, even in the initial iterations of BP decoding where the LLR values are not evolved yet. This is not the case for CRC-polar concatenated code since CRC is connected to all the information bits of polar code, some of which are highly unreliable during the early iterations of BP decoding.

	In order to address the above issue, we first run BP decoding on the polar code factor graph for a maximum of $I_{\text{thr}}$ iterations and if the BP decoding is not successful after $I_{\text{thr}}$ iterations, we then continue the BP decoder on the CRC-polar concatenated factor graph. In order to determine if the decoder has succeeded, we use the CRC at each iteration as an early stopping criterion. The proposed decoding algorithm is summarized in Algorithm~\ref{algCPBP}. The LLR vectors $\bm{l}^i_s$ and $\bm{r}^i_s$ at all stages and iterations are initialized as explained in Section \ref{sec:polar:BPD}. The \texttt{BP\_PolarLeft} and \texttt{BP\_PolarRight} functions compute (\ref{PolarPE_left}) and (\ref{PolarPE_right}) at all the bit indices to perform the polar right-to-left and left-to-right LLR updates, respectively. The estimated message word $\bm{\hat{u}}$ is obtained at every iteration by making a hard decision based on $\bm{l}^i_0$ and $\bm{r}^i_0$, which is done by executing (\ref{hardDec}) in the \texttt{HardDecision} function. A CRC is then applied on $\bm{\hat{u}}$ and the decoding can be early terminated if the CRC is satisfied. After $I_{\text{thr}}$ iterations, if the decoding is not terminated, BP decoding on the CRC-polar factor graph is carried out in the \texttt{BP\_CRC} function. It is worth mentioning that BP decoding after $I_{\text{thr}}$ iterations runs on the concatenated CRC-polar factor graph at every iteration.
	
	\begin{algorithm}
		\caption{CPBP Decoding Algorithm}
		\label{algCPBP}
		\DontPrintSemicolon
		\SetKwInOut{Input}{Input}
		\SetKwInOut{Output}{Output}
		\SetKwFunction{BPPolarLeft}{BP\_PolarLeft}
		\SetKwFunction{BPPolarRight}{BP\_PolarRight}
		\SetKwFunction{BPPolarCRC}{BP\_CRC}
		\SetKwFunction{CRCCheck}{CRC\_IsTerminated}
		\SetKwFunction{HardDecision}{HardDecision}
		\SetKwFunction{Terminate}{Terminate}
		\SetKwFunction{Init}{Init}
		
		\Input{$I_{\max}, I_{\text{thr}},n$} 
		\Output{$\bm{\hat{u}}$}
		Initialize $\bm{l}^i_s,\bm{r}^i_s$ ($1 \leq i \leq I_{max}$, $0 \leq s \leq n$)\\
		\For{$i \leftarrow 1$ \KwTo $I_{\max}$}{
			\For{$s \leftarrow n-1$ \KwTo $0$}{
				$\bm{l}^i_s \leftarrow$ \BPPolarLeft{$\bm{l}^i_{s+1},\bm{r}^{i-1}_s$} \\	
			}	
			\If{$i > I_{\textup{thr}}$}{
				$\bm{r}^i_0 \leftarrow$ \BPPolarCRC{$\bm{l}^i_0$} \\ 
			}
			$\bm{\hat{u}} \leftarrow $ \HardDecision{$\bm{r}^i_0 + \bm{l}^i_0$} \\
			\If{$\bm{\hat{u}}$ satisfies CRC}{
				Terminate
			}		
			\If{$i \le I_{\max} - 1$}{
				\For{$s \leftarrow 1$ \KwTo $n-1$}{
					$\bm{r}^i_s \leftarrow$ \BPPolarRight{$\bm{l}^i_s,\bm{r}^i_{s-1}$} \\
				}
			}
		}
		\Return{$\bm{\hat{u}}$}
	\end{algorithm}
	
\begin{figure}[t]
		\vspace*{1\baselineskip}
		\centering
		\input{Figures/fer_CPBP_N128_K80_CRC16.tikz}		
		\hspace{30pt} \ref{perf-legend-CPBP}
		\caption{FER performance of CPBP decoding for $\mathcal{P}(128,80)$ and a $16$-bit CRC used in 5G.}
		\label{fig:CPBP_per}	
		\vspace*{-0.6\baselineskip}
	\end{figure}
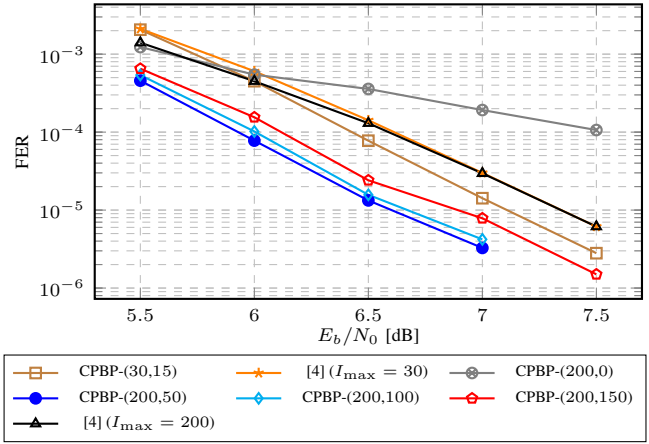

	Fig.~\ref{fig:CPBP_per} shows the FER performance of the proposed CPBP algorithm in comparison with the CRC-aided BP decoder of \cite{RenBPTerminate}, for the $\mathcal{P}(128,80)$ concatenated with the $16$-bit CRC, which is selected for 5G \cite{3gpp_report}. In this figure, we set $I_{\max} \in \{30,200\}$ and we set $I_{\text{thr}} \in \{15,30\}$ when $I_{\max} = 30$ and $I_{\text{thr}} \in \{0,50,100,150,200\}$ when $I_{\max} = 200$. We denote CPBP decoding with parameters $I_{\max}$ and $I_{\text{thr}}$ as CPBP-($I_{\max}$,$I_{\text{thr}}$). Note that CPBP-($I_{\max}$,$I_{\max}$) is equivalent to the decoder in \cite{RenBPTerminate} and CPBP-($I_{\max}$,$0$) is the direct application of the approach in \cite{Guo}. It can be seen that, CPBP-($30$,$15$) provides a gain of almost $0.25$~dB in comparison with the CRC-aided BP decoder of \cite{RenBPTerminate} at the target FER of $10^{-5}$. In addition, among the selected $I_{\text{thr}}$ for $I_{\max} = 200$, CPBP-($200$,$50$) provides the best error-correction performance at the target FER of $10^{-5}$. Furthermore, CPBP-($200,50$) has an error-correction performance gain of about $0.75$~dB at FER~$=10^{-5}$ in comparison with the CRC-aided BP decoder of \cite{RenBPTerminate}. It is worth mentioning that increasing $I_{\max}$ does not improve the error probabilities of the conventional BP decoder in \cite{RenBPTerminate} at high $E_b/N_0$ regime. On the contrary, the FER of the proposed CPBP decoder is greatly benefited from a high value of $I_{\max}$ as observed from Fig.~\ref{fig:CPBP_per}.
	
	We now evaluate the latency of the proposed CPBP decoding scheme and compare it with state-of-the-art. The latency of a BP-based decoder can be measured using the number of time steps required to finish the decoding process \cite{arikan2010polar}. Let us consider the decoding process terminates at iteration $I_\text{ET}$ ($1 \le I_{\text{ET}} \le I_{\max}$). Then the decoding latency of a conventional BP decoder with early stopping criterion can be represented as
	\begin{equation}
	\mathcal{T}_{\text{BP}} = (2n-1)(I_{\text{ET}}-1)+n \text{.}
	\label{eq:BPLatency}
	\end{equation}
The latency of the proposed CPBP decoder depends on when the decoding process terminates and can be represented as
	\begin{equation}
	\mathcal{T}_{\text{CPBP}} \!=\! 
	\begin{cases}
	(2n\!-\!1)(I_{\text{ET}}\!-\!1)\!+\!n, & \!\!\!\text{ if } I_{\text{ET}} \!\le\! I_{\text{thr}},\\
	(2n\!-\!1)(I_{\text{ET}}\!-\!1)\!+\!n\!+\!2(I_{\text{ET}}\!-\!I_{\text{thr}}),  & \!\!\!\text{ otherwise.}
	\end{cases}
	\label{eq:CPBPLatency}
	\end{equation}
	In fact, if $I_{\text{ET}} \le I_{\text{thr}}$, (\ref{eq:CPBPLatency}) reverts to (\ref{eq:BPLatency}) since the CPBP decoder terminates without traversing the CRC factor graph. It should be noted that the worst case latency of the BP decoder and the proposed CPBP decoder can be calculated using (\ref{eq:BPLatency}) and (\ref{eq:CPBPLatency}) respectively, by setting $I_{\text{ET}} = I_{\max}$.

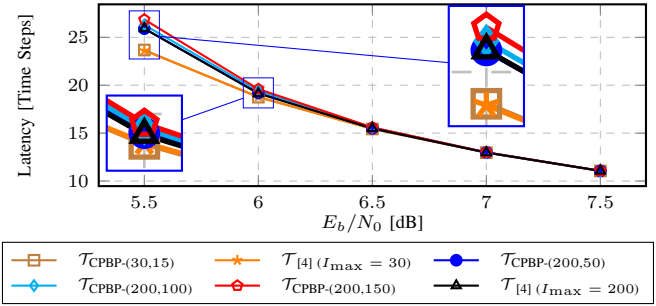
\begin{figure}[t]
		\vspace*{1\baselineskip}
		\centering
		\input{Figures/timesteps_CPBP_N128_K80_CRC16.tikz}		
		\hspace{30pt} \ref{latency-legend-CPBP}
		\caption{Average decoding latency of CPBP decoding for $\mathcal{P}(128,80)$ and a $16$-bit CRC used in 5G.}
		\label{fig:CPBP_latency}	
		\vspace*{-0.6\baselineskip}		
	\end{figure}

	Fig.~\ref{fig:CPBP_latency} illustrates the average latency of the proposed CPBP decoding algorithm in comparison with a conventional CRC-aided BP decoder of \cite{RenBPTerminate} for the same code as in Fig.~\ref{fig:CPBP_per}. For the proposed decoders, we set $I_{\text{thr}} \in \{15,30\}$ for $I_{\max}=30$, and $I_{\text{thr}} \in \{50,100,150,200\}$ for $I_{\max}=200$. It can be seen that the proposed CPBP algorithm incurs negligible latency overhead in comparison with \cite{RenBPTerminate}, while providing significant performance gain. Moreover, the average latency of the CPBP decoder when $I_{\max} = 30$ is always smaller than that of the CPBP decoder when $I_{\max} = 200$. This average latency saving is more significant for lower $E_b/N_0$ values. Furthermore, the worst case latency of CPBP($200$,$50$) is $2887$ time steps, and that of CPBP($30$,$15$) is $407$ time steps which is only $14\%$ of the worst case latency of CPBP($200$,$50$). For applications with stringent latency requirements, a small $I_{\max}$ is needed. However, the latency saving as a result of a small $I_{\max}$ comes at the cost of error-correction performance loss as shown in Fig.~\ref{fig:CPBP_per}. In the next section, we propose a method to improve the error-correction performance of CPBP decoding for small values of $I_{\max}$, by using trainable weights.
	
\section{Neural CRC-Polar BP Decoding}
\label{sec:NCPBP}
		
	\begin{figure*}[th]
		\centering
		\includegraphics[width=0.77\linewidth]{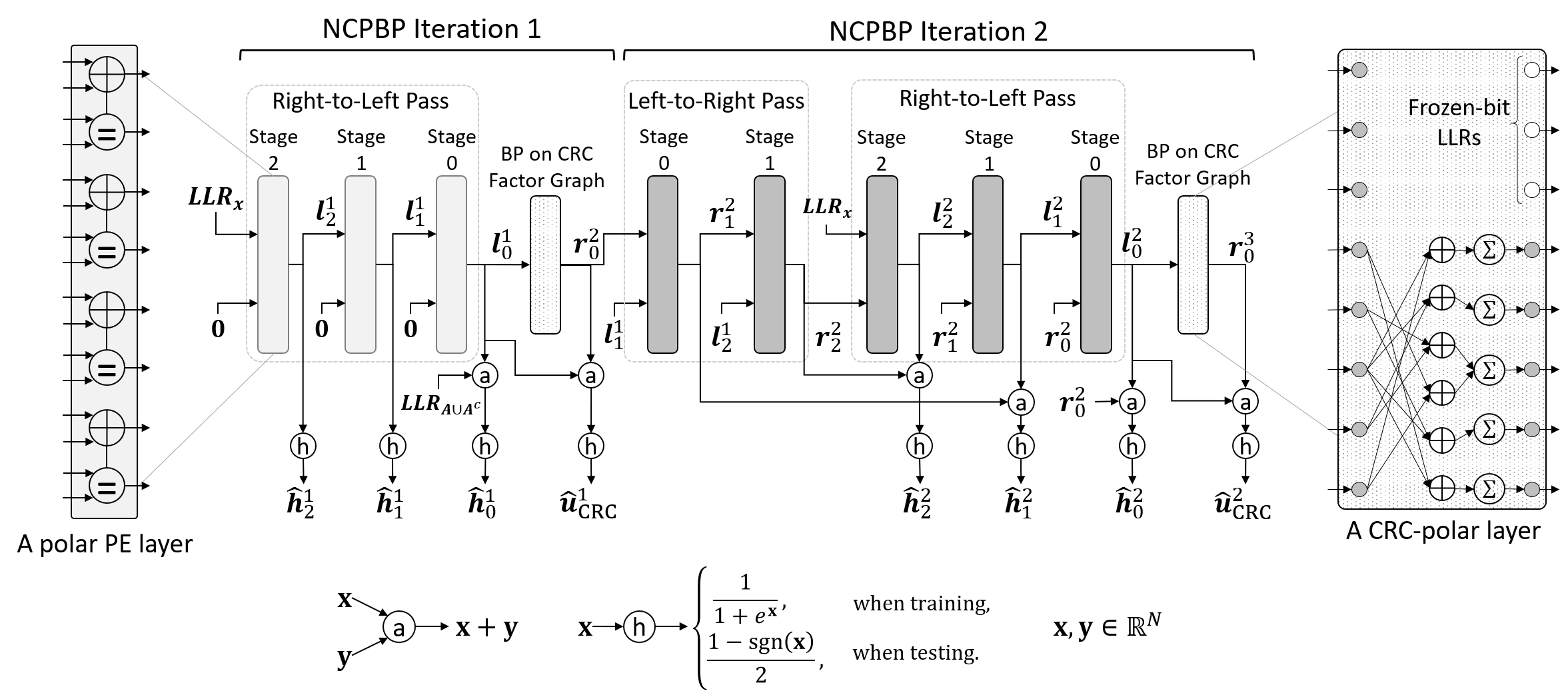}
		\caption{NCPBP architecture with $I_{\max}=2$ and $I_{\text{thr}}=0$ for $\mathcal{P}(8,3)$ concatenated with a $2$-bit CRC.}
		\label{fig:NCPBPArch}
		\vspace*{-0.8\baselineskip}		
	\end{figure*}
	
	In this section, we propose the NCPBP decoder to improve the error-correction performance of CPBP decoding. The NCPBP decoder assigns trainable weights to the edges of the CRC-polar concatenated factor graph. Therefore, the NCPBP decoder resembles a neural network architecture by mapping the message updates of CPBP decoding to different computational layers in the neural network. In other words, each computational layer of the neural network is represented either as a set of PEs for BP decoding on the factor graph of polar codes, or as a set of operations required to perform BP decoding on the CRC factor graph. This network architecture greatly simplifies the training process since it can be adapted to recent deep learning frameworks, e.g. Tensorflow \cite{TensorFlow}.
	
	Fig.~\ref{fig:NCPBPArch} depicts the architecture of the proposed NCPBP decoder for $\mathcal{P}(8,3)$, with $I_{\max}=2$ and $I_{\text{thr}}=0$. The architecture contains the unrolled CRC-polar concatenated factor graph. Therefore, the message updates of BP decoding on polar codes at a computational layer is represented as the ones in Fig.~\ref{figBPDec:c} and Fig.~\ref{figBPDec:d}. In order to assign the weights to the $\frac{N}{2}$ parallel PEs at polar code computational layers, we represent the product of the weights $w_3$ and $w_4$ in (\ref{PE_NNMSRNN_left}), and the product of the weights $w_9$ and $w_{10}$ in (\ref{PE_NNMSRNN_right}), as single trainable weights $w_{3,4}$ and $w_{9,10}$, respectively. This is due to the fact that the product of two trainable weights can be merged into one as the new weight can also be optimized during training. In addition, we merge the weights $w_1$ and $w_2$ in (\ref{PE_NNMSRNN_left}) into $w_{1,2}$, and the weights $w_7$ and $w_8$ in (\ref{PE_NNMSRNN_right}) into $w_{7,8}$, to further reduce the number of trainable weights. As a result, we define the weight assignment scheme of a PE in NCPBP decoding as
	\begin{align}
	\label{PE_NNBP_left}
	\begin{split}
	\begin{cases}
	l^{i}_{t,s} &= w_0 \tilde{f}(l^{i}_{t,k},w_{1,2} (r^{i}_{j,s} + l^{i}_{j,k}))\text{,}\\
	l^{i}_{j,s} &= w_{3,4}\tilde{f}(l^{i}_{t,k},r^{i}_{t,s}) + w_5 l^{i}_{j,k}\text{,}\\
	\end{cases}
	\end{split}
	\end{align}
	\begin{align}
	\label{PE_NNBP_right}
	\begin{split}
	\begin{cases}
	r^{i}_{t,k} &= w_6 \tilde{f}(r^{i}_{t,s},w_{7,8} (l^{i-1}_{j,k} + r^{i}_{j,s}))\text{,}\\
	r^{i}_{j,k} &= w_{9,10} \tilde{f}(r^{i}_{t,s},l^{i-1}_{t,k}) + w_{11} r^{i}_{j,s}\text{.}
	\end{cases}
	\end{split}
	\end{align}
	For the BP decoding on the CRC factor graph of the proposed NCPBP decoder, we adopt the weight assignment scheme of the NNMS-RNN decoder in \cite{Nachmani_STSP}. It should be noted that the polar code computational layers share the same set of weights in each iteration of the proposed NCPBP decoding, while this set of weights is different for different iterations. This is illustrated in Fig.~\ref{fig:NCPBPArch}, in which the layers depicted in the same color indicate that they use the same set of weights. On the contrary, the weights used in all the CRC layers are shared among all the decoding iterations of NCPBP. This is particularly useful in order to limit the number of required weights for NCPBP.
		
	The NCPBP decoding algorithm starts by a right-to-left message update at iteration $1$. At the $i$-th iteration and the $s$-th stage of the NCPBP decoder, $\bm{l}^i_s$ and $\bm{r}^i_s$ denote the LLR vectors of the right-to-left and left-to-right message updates computed by a polar code PE layer, respectively. Furthermore, the output LLR vector of the CRC layer is denoted as $\bm{r}^{i+1}_0$. The hard estimated values of all the stages in the polar code factor graph are obtained at the right-to-left message updates, denoted as $\bm{\hat{h}}^i_s$, while the hard estimated values derived from the CRC layer is denoted as $\bm{\hat{u}}^i_{\text{CRC}}$.
	
	The weights of all the polar code and CRC computational layers are trained using a multiloss function defined as
	\begin{equation}
		L = \sum_{i=1}^{I_{\max}}\sum_{s=0}^{n-1}H_{\text{CE}}(\bm{\hat{h}}^i_s,\bm{h_{s}})+\sum_{i'=I_{\text{thr}}+1}^{I_{\max}} H_{\text{CE}}(\bm{\hat{u}}^{i'}_{\text{CRC}},\bm{u}) \text{,}
	\end{equation}
	where $H_{\text{CE}}$ is the cross-entropy function, and $\bm{h_{s}}$ is the correct hard value vector at stage $s$ of the polar code factor graph which is obtained from the training samples. Note that in the testing phase, only the hard estimated values at stage $0$ of the polar code factor graph, i.e. $\bm{\hat{h}}^i_0$ $(1 \leq i \leq I_{\max})$, and the hard estimated values at the CRC layer, i.e. $\bm{\hat{u}}^i_{\text{CRC}}$, $(I_{\text{thr}} < i \leq I_{\max})$, are required to obtain the decoded message bits.
	
	We evaluate the proposed NCPBP decoder for $\mathcal{P}(128,80)$ concatenated with a $16$-bit CRC which is also used in Section~\ref{sec:CPBP}, and we compare the error-correction performance and latency of NCPBP with those of \cite{RenBPTerminate,Xu,Nachmani_STSP}. All the neural BP-based decoders in this section are trained using stochastic gradient descent with RMSPROP optimizer \cite{hinton2012neural} and the learning rate is set to $0.001$. We use Tensorflow \cite{TensorFlow} as our deep learning framework. Since all the considered neural BP-based decoders satisfy the symmetry conditions \cite{Richardson_ITIT_2001}, we collect $100,000$ zero codewords at each $E_b/N_0$ value for training, where $E_b/N_0 \in \{4,4.5,5,5.5\}$~dB. All the weights of all the neural BP-based decoders are initialized to one and all the LLR values are clipped to be in the interval of $[-20,20]$. The mini-batch size is set to $64$ and each neural decoder is trained for $40$ epochs. To evaluate the error-correction performance, randomly generated codewords are used during the testing phase and each decoder simulates at least $10,000$ codewords until it obtains at least $50$ frames in error.
	
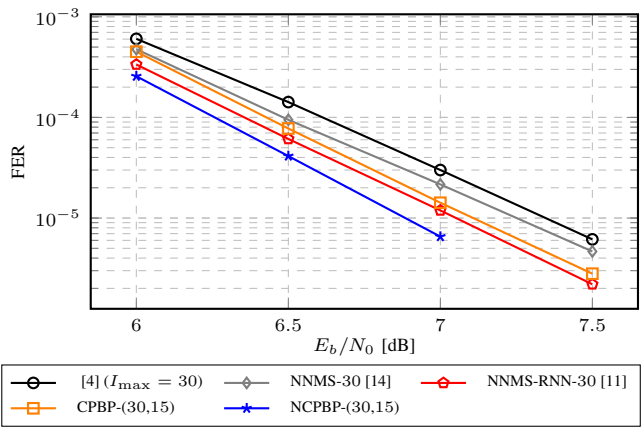
\begin{figure}[t]
		\centering
		\input{Figures/fer_NCPBP_N128_K80_CRC16.tikz}		
		\hspace{30pt}	\ref{perf-legend-NCPBP}
		\caption{FER performance of NCPBP decoding for $\mathcal{P}(128,80)$ and a $16$-bit CRC used in 5G.}
		\label{fig:NCPBP}
	\end{figure}

	Fig.~\ref{fig:NCPBP} compares the error-correction performance of the proposed NCPBP decoder with state-of-the-art BP-based decoders in \cite{RenBPTerminate,Xu,Nachmani_STSP}. We use the NNMS-$I_{\max}$ decoder of \cite{Xu} and the NNMS-RNN-$I_{\max}$ decoder of \cite{Nachmani_STSP} for our comparisons. In all the decoders, we set $I_{\max} = 30$. At a target FER of $10^{-5}$, the proposed NCPBP decoder provides about $0.5$~dB gain with respect to \cite{RenBPTerminate}, $0.4$~dB gain with respect to \cite{Xu}, and $0.2$~dB gain with respect to \cite{Nachmani_STSP}. Compared to the CPBP decoder of Section~\ref{sec:CPBP}, the proposed NCPBP provides $0.25$~dB FER performance improvement.
	
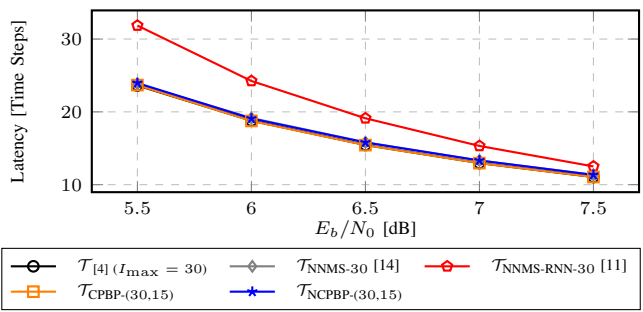
\begin{figure}[t]
		\centering
		\input{Figures/timesteps_NCPBP_N128_K80_CRC16.tikz}		
		\hspace{30pt}	\ref{latency-legend-NCPBP}
		\caption{Average decoding latency of NCPBP decoding for $\mathcal{P}(128,80)$ and a $16$-bit CRC used in 5G.}
		\label{fig:NCPBP_latency}
		\vspace*{-0.5\baselineskip}
	\end{figure}

	Fig.~\ref{fig:NCPBP_latency} illustrates the average latency requirements of the NCPBP decoder compared to the state-of-the-art decoders in \cite{RenBPTerminate,Xu,Nachmani_STSP}. It can be seen that while the average latency of the NCPBP decoder is similar to that of the decoders in \cite{RenBPTerminate,Xu}, it is always better than that of \cite{Nachmani_STSP}. In addition, NCPBP incurs almost no latency overhead with respect to the proposed CPBP decoder while having a notably smaller error probability.
	
	Table~\ref{tab:param} shows the number of weights required for the proposed NCPBP decoder in comparison with the neural BP decoders of \cite{Xu,Nachmani_STSP}. The proposed NCPBP decoder requires $28\%$ fewer weights with respect to the decoder in \cite{Nachmani_STSP}. The decoder in \cite{Xu} requires $46\%$ of the weights that is required by the proposed NCPBP decoder. However, the smaller number of weights in \cite{Xu} results in significant error-correction performance loss as shown in Fig.~\ref{fig:NCPBP}.
	
	\begin{table}[t]
		\vspace*{1\baselineskip}
		\centering
		\caption{Number of weights required by different neural BP decoders.}
		\setlength{\extrarowheight}{2.5pt}
		\begin{tabular}{lc}
			\toprule
			Decoder & Number of weights \\  
			\midrule
			NNMS-$30$ \cite{Xu} & $3840$\\
			NNMS-RNN-$30$ \cite{Nachmani_STSP} & $11520$ \\
			NCPBP-($30$,$15$) & $8288$ \\
			\bottomrule
		\end{tabular}
		\label{tab:param}
		\vspace*{-0.5\baselineskip}
	\end{table}
	
	\section{Conclusion} 
	\label{sec:conclude}
	
	In this paper, we first proposed a cyclic redundancy check (CRC)-polar belief propagation (BP) (CPBP) decoding algorithm which exploits concatenated factor graphs of polar codes and CRC, by passing the extrinsic information between the two factor graphs. We showed that the proposed CPBP decoding brings significant error-correction performance improvements in comparison with a conventional BP decoder when a large maximum number of BP iterations is used. We further proposed a neural CPBP (NCPBP) decoding algorithm which further improves the error probabilities of CPBP by assigning trainable weights to the edges of the CRC-polar concatenated factor graph. We showed that the NCPBP decoding algorithm can be used in applications which require stringent latency requirements and that it can benefit from the CRC which is present in 5G. Our results for a polar code of length $128$ with $80$ information bits concatenated with a CRC of length $16$ show that the proposed NCPBP decoding algorithm obtains up to $0.4$~dB error-correction performance improvement with respect to the state of the art, at a target frame error rate of $10^{-5}$, while incurring negligible latency overhead.
	


\end{document}

%% file: Figures/fer_CPBP_N128_K80_CRC16.tikz
\begin{tikzpicture}[spy using outlines = {rectangle, magnification=2.5, connect spies}]
  \pgfplotsset{	
    label style = {font=\fontsize{7pt}{7}\selectfont},
    tick label style = {font=\fontsize{7pt}{7}\selectfont}
  }

\begin{axis}[
	scale = 1,
    ymode=log,
    xlabel={$E_b/N_0$ [\text{dB}]}, xlabel style={yshift=0.8em},
    ylabel={FER}, ylabel style={yshift=-0.75em},
    grid=both,
    legend cell align={left},
    ymajorgrids=true,
    xmajorgrids=true,
    grid style=dashed,
    width=1\columnwidth, height=5.5cm,
    thick,
    mark size=2,
    legend style={
      column sep= 2mm,
      font=\fontsize{6pt}{7.2}\selectfont,
    },
    legend to name=perf-legend-CPBP,
	legend columns=3,
]

\addplot[
color=brown,
mark=square,
thick,
mark options={solid},
mark size=2,
]
table {
	5.5	2.06E-03
	6	4.50E-04
	6.5	7.76E-05
	7	1.42E-05
	7.5	2.80E-06
};
\addlegendentry{CPBP-($30$,$15$)}

\addplot[
color=orange,
mark=star,
thick,
mark options={solid},
mark size=2,
]
table {
	5.5	0.00210801
	6	0.000603865
	6.5	0.000142008
	7	3.00E-05
	7.5	6.14E-06	
};
\addlegendentry{\cite{RenBPTerminate} ($I_{\max}=30$)}

\addplot[
color=gray,
mark=otimes,
thick,
mark size=2,
]
table {
5.5	0.00123177
6	0.000545834
6.5	0.000357964
7	0.000191907
7.5	0.000106326
};
\addlegendentry{CPBP-($200$,$0$)}

\addplot[
    color=blue,
    mark=*,
    thick,
    mark size=2,
]
table {
5.5	4.57E-04
6	7.78E-05
6.5	1.33E-05
7	3.27E-06
};
\addlegendentry{CPBP-($200$,$50$)}

\addplot[
    color=cyan,
    mark=diamond,
    thick,
    mark size=2,
]
table {
5.5	5.39E-04
6	1.02E-04
6.5	1.57E-05
7	4.21E-06	
};
\addlegendentry{CPBP-($200$,$100$)}

\addplot[
color=red,
mark=pentagon,
thick,
mark size=2,
]
table {
5.5	6.51E-04
6	1.55E-04
6.5	2.41E-05
7	7.85E-06
7.5	1.50E-06
};
\addlegendentry{CPBP-($200$,$150$)}

\addplot[
color=black,
mark=triangle,
thick,
mark size=2,
]
table {
5.5	0.0014028
6	0.000449745
6.5	1.30E-04
7	2.96E-05
7.5	6.14E-06
};
\addlegendentry{\cite{RenBPTerminate} ($I_{\max}=200$)}

\end{axis}
\end{tikzpicture}

%% file: Figures/timesteps_CPBP_N128_K80_CRC16.tikz
\begin{tikzpicture}[spy using outlines = {rectangle, magnification=2.5, connect spies}]
\pgfplotsset{	
	label style = {font=\fontsize{7pt}{7}\selectfont},
	tick label style = {font=\fontsize{7pt}{7}\selectfont}
}

\begin{axis}[
scale = 1,
xlabel={$E_b/N_0$ [\text{dB}]}, xlabel style={yshift=0.8em},
ylabel={Latency [Time Steps]}, ylabel style={yshift=-0.75em},
grid=both,
legend cell align={left},
ymajorgrids=true,
xmajorgrids=true,
grid style=dashed,
width=1\columnwidth, height=4cm,
thick,
mark size=2,
legend style={
	column sep= 2mm,
	font=\fontsize{6pt}{7.2}\selectfont,
},
    legend to name=latency-legend-CPBP,
	legend columns=3,
]

\addplot[
color=brown,
mark=square,
thick,
mark size=2,
]
table {
	5.5	23.6758
	6	18.7521
	6.5	15.4331
	7	12.96
	7.5	11.051072
};
\addlegendentry{$\mathcal{T}_{\text{CPBP-($30$,$15$)}}$}

\addplot[
color=orange,
mark=star,
thick,
mark size=2,
]
table {
	5.5	23.62
	6	18.79
	6.5	15.45
	7	12.97
	7.5	11.05
};
\addlegendentry{$\mathcal{T}_{\text{\cite{RenBPTerminate} ($I_{\max}=30$)}}$}

\addplot[
color=blue,
mark=*,
thick,
mark size=2,
]
table {
5.5	25.9016
6	1.92E+01
6.5	1.55E+01
7	12.9822
7.5	11.060869	
};
\addlegendentry{$\mathcal{T}_{\text{CPBP-($200$,$50$)}}$}

\addplot[
color=cyan,
mark=diamond,
thick,
mark size=2,
]
table {
5.5	26.2917
6	19.405
6.5	15.5862
7	12.9948
7.5	11.056418
};
\addlegendentry{$\mathcal{T}_{\text{CPBP-($200$,$100$)}}$}

\addplot[
color=red,
mark=pentagon,
thick,
mark size=2,
]
table {
5.5	26.8618
6	1.96E+01
6.5	1.56E+01
7	1.30E+01
7.5	11.065088
};
\addlegendentry{$\mathcal{T}_{\text{CPBP-($200$,$150$)}}$}

\addplot[
color=black,
mark=triangle,
thick,
mark size=2,
]
table {
5.5	25.9145
6	19.1363
6.5	15.4937
7	12.9738
7.5	11.061139	
};
\addlegendentry{$\mathcal{T}_{\text{\cite{RenBPTerminate} ($I_{\max}=200$)}}$}

\coordinate (spypoint1) at (axis cs:6,19.2);
\coordinate (magnifyglass1) at (axis cs:5.5,15);  

\coordinate (spypoint2) at (axis cs:5.5,25.25);
\coordinate (magnifyglass2) at (axis cs:7,22);  

\end{axis}
\spy [blue, width=1cm, height=1cm] on (spypoint1) in node[fill=white] at (magnifyglass1);
\spy [blue, width=1cm, height=1.6cm] on (spypoint2) in node[fill=white] at (magnifyglass2);
\end{tikzpicture}

%% file: Figures/fer_NCPBP_N128_K80_CRC16.tikz
\begin{tikzpicture}[spy using outlines = {rectangle, magnification=2, connect spies}]
\pgfplotsset{	
	label style = {font=\fontsize{7pt}{7}\selectfont},
	tick label style = {font=\fontsize{7pt}{7}\selectfont}
}

\begin{axis}[
scale = 1,
ymode=log,
xlabel={$E_b/N_0$ [\text{dB}]}, xlabel style={yshift=0.8em},
ylabel={FER}, ylabel style={yshift=-0.75em},
xtick={5.5,6,6.5,7,7.5},
grid=both,
ymajorgrids=true,
xmajorgrids=true,
grid style=dashed,
width=1\columnwidth, height=5.5cm,
thick,
mark size=2,
legend cell align={left},
legend style={
	column sep= 2mm,
	font=\fontsize{6pt}{7.2}\selectfont,
},
  legend to name=perf-legend-NCPBP,
  legend columns=3,
]

\addplot[
color=black,
mark=o,
thick,
mark size=2,
]
table {
	6	0.000603865
	6.5	0.000142008
	7	3.00E-05
	7.5	6.14E-06
};
\addlegendentry{\cite{RenBPTerminate} ($I_{\max}=30$)}

\addplot[
color=gray,
mark=diamond,
thick,
mark size=2,
]
table {
	6	0.00047619
	6.5	9.47E-05
	7	2.16E-05
	7.5	4.6562E-06
};
\addlegendentry{NNMS-$30$ \cite{Xu}}

\addplot[
color=red,
mark=pentagon,
thick,
mark size=2,
]
table {
	6	0.00033557
	6.5	6.11E-05
	7	0.00001191
	7.5	2.1978E-06
};
\addlegendentry{NNMS-RNN-$30$ \cite{Nachmani_STSP}}



\addplot[
color=orange,
mark=square,
thick,
mark size=2,
]
table {
	6	4.50E-04
	6.5	7.76E-05
	7	1.42E-05
	7.5	2.80E-06
};
\addlegendentry{CPBP-($30$,$15$)}

\addplot[
color=blue,
mark=star,
thick,
mark size=2,
]
table {
	6	0.00025641
	6.5	4.13E-05
	7	6.5377E-06
};
\addlegendentry{NCPBP-($30$,$15$)}


\end{axis}
\end{tikzpicture}

%% file: Figures/timesteps_NCPBP_N128_K80_CRC16.tikz
\begin{tikzpicture}[spy using outlines = {rectangle, magnification=2.5, connect spies}]
\pgfplotsset{	
	label style = {font=\fontsize{7pt}{7}\selectfont},
	tick label style = {font=\fontsize{7pt}{7}\selectfont}
}

\begin{axis}[
scale = 1,
xlabel={$E_b/N_0$ [\text{dB}]}, xlabel style={yshift=0.8em},
ylabel={Latency [Time Steps]}, ylabel style={yshift=-0.75em},
xtick={5.5,6,6.5,7,7.5},
grid=both,
legend cell align={left},
ymajorgrids=true,
xmajorgrids=true,
grid style=dashed,
width=1\columnwidth, height=4cm,
thick,
mark size=2,
legend style={
	column sep= 2mm,
	font=\fontsize{6pt}{7.2}\selectfont,
},
    legend to name=latency-legend-NCPBP,
	legend columns=3,
]

\addplot[
color=black,
mark=o,
thick,
mark size=2,
]
table {
	5.5	23.62
	6	18.79
	6.5	15.45
	7	12.97
	7.5	11.05
};
\addlegendentry{$\mathcal{T}_{\text{\cite{RenBPTerminate} ($I_{\max}=30$)}}$}

\addplot[
color=gray,
mark=diamond,
thick,
mark size=2,
]
table {
	5.5	23.84575862
	6	19.1576
	6.5	15.82687689
	7	13.3306566
	7.5	11.3598083
};
\addlegendentry{$\mathcal{T}_{\text{NNMS-}30}$ \cite{Xu}}

\addplot[
color=red,
mark=pentagon,
thick,
mark size=2,
]
table {
	5.5	31.85442424
	6	24.24192617
	6.5	19.11785941
	7	15.3195703
	7.5	12.50785416
};
\addlegendentry{$\mathcal{T}_{\text{NNMS-RNN-}30}$ \cite{Nachmani_STSP}}



\addplot[
color=orange,
mark=square,
thick,
mark size=2,
]
table {
	5.5	23.6758
	6	18.7521
	6.5	15.4331
	7	12.96
	7.5	11.051072
};
\addlegendentry{$\mathcal{T}_{\text{CPBP-($30$,$15$)}}$}

\addplot[
color=blue,
mark=star,
thick,
mark size=2,
]
table {
	5.5	23.95655556
	6	19.08363636
	6.5	15.79582353
	7	13.3188326
	7.5	11.34
};
\addlegendentry{$\mathcal{T}_{\text{NCPBP-($30$,$15$)}}$}

\end{axis}
\end{tikzpicture}